# Synthesis and optical properties of large-scale single-crystalline two-dimensional semiconductor WS$_2$ monolayer from chemical vapor deposition

Chunxiao Cong[1], Jingzhi Shang[1], Xing Wu[2], Bingchen Cao[1], Namphung Peimyoo[1], Caiyu Qiu[1], Litao Sun[2], and Ting Yu[1, 3, 4, *]

[1] Division of Physics and Applied Physics, School of Physical and Mathematical Sciences, Nanyang Technological University, 637371, Singapore

[2] SEU-FEI Nano-Pico Center, Key Laboratory of MEMS of Ministry of Education, School of Electrical Science and Engineering, Southeast University, Nanjing 210096, China

[3] Department of Physics, Faculty of Science, National University of Singapore, 117542, Singapore

[4] Graphene Research Center, National University of Singapore, 117546, Singapore

[*]Address correspondence to yuting@ntu.edu.sg

**Two-dimensional (2D) transition metal dichalcogenides (TMDs), especially MoS$_2$ and WS$_2$ recently attract extensive attentions due to their rich physics and great potential applications. Superior to graphene, MS$_2$ (M = Mo/W) monolayers have a native direct energy gap in visible frequency range. This promises great future of MS$_2$ for optoelectronics. To exploit properties and further develop more applications, producing large-scale single crystals of MS$_2$ by a facile method is highly demanded. Here, we report the synthesis of large-scale triangular single crystals of WS$_2$ monolayer from a chemical vapor deposition process and systematic optical studies of such WS$_2$ monolayers. The observations of high yield of light emission and valley-selective circular dichroism experimentally evidence the high optical quality of the WS$_2$ monolayers. This work paves the road to fabricate large-scale single crystalline 2D semiconductors and study their fundamentals. It must be very meaningful for exploiting great potentials of WS$_2$ for future optoelectronics.**

Being entangled in controlling the electronic properties of graphene for next-generation electronics[1-2], monolayer transition metal dichalcogenides such as MS$_2$ (M = Mo, W) are intriguing great interest as two-dimensional (2D) semiconductors with a native direct energy gap in visible frequency range[3-4]. There are comprehensive and intensive studies on monolayer $MoS_2$ including optical and electronic properties[5-9], valleytronics[10-13], strain effects[14-16], thermal effects[17], and so on. However, the investigation on $WS_2$ has just started. Similar to 2H-$MoS_2$, monolayer 2H-$WS_2$ can be constructed by sandwiching two atomic layers of S and one atomic layer of W through covalent bonds of W-S, where W locates at the body center of a trigonal-prismatic case formed by six S atoms. Confinement of charge carriers inside the horizontal atomic plane gradually enlarges energy gaps when thinning $WS_2$ layers[18]. Instead of an indirect energy gap for multiple layers, a direct energy gap of ~2 eV at the corners (K and K' points) of the Brillion Zone could be formed in the monolayer $WS_2$ as clearly demonstrated by both theoretical and experimental studies[5,19-21]. The immediate consequence, also a benefit of the existence of such direct band gap is the significant enhancement of visible light emission. In monolayer $WS_2$, breaking inversion symmetry leads to the strong spin-orbit coupling and the splitting of valence bands at K/K' point with a sub-gap of around 0.4 eV[22]. Furthermore, the split spins at the time reversed K and K' valleys have the opposite signs. Thus, such spin-valley coupling offers extra degree of freedom to charge carriers in $WS_2$ monolayer. Though it has not been reported in monolayer $WS_2$, as predicted by the theory and observed by the experiments in monolayer $MoS_2$, the non-equilibrium charge carrier imbalance at two valleys has been well revealed by the remarkable difference of absorption of left- (σ-) and right-handed (σ+) circular polarized lights at two valleys[5,10-13,23]. All these interesting and important properties, plus the newly revealed potential in the flexible heterostructures of graphene-$WS_2$ stacks[24-25] guarantee a promising future of $WS_2$ as the candidate of next generation of nanoelectronics, spintronics, valleytronics and optoelectronics. However, compared to graphene, it is very difficult to prepare $MS_2$ monolayers, and atomically thin $MS_2$ flakes made by mechanical exfoliation are much smaller, in fact too small to be well characterized and processed for devices. Most recently, chemical vapor deposition (CVD) has been used to successfully grow large-scale single crystals of monolayer $MoS_2$[7,26-29]. However, the closed count partner - $WS_2$ single crystalline monolayers, can only be synthesized by CVD with dimension of a few microns. In this article, we report our study of growing large-scale triangular single crystals of WS$_2$ monolayer up to hundreds of microns with high optical quality by one step of direct sulfurization of WO$_3$ powders and probing their optical properties.

**Results**

**Growth and characterization of large-scale single crystalline WS$_2$ monolayers.** As illustrated in Fig. 1a, the growth process is very different with the previous reported CVD growth of WS$_2$ thin flakes[30]. Without the pre-deposition of WO$_3$ thin film in a high vacuum chamber, in this work, rather than by two-step process, we fabricate large-scale single crystalline WS$_2$ monolayers by one step of direct sulfurization of WO$_3$ powders by sulfur powders at 750 °C (the detailed growth mechanism is proposed below). The substrates used here are well cleaned 300 nm SiO$_2$/Si wafer. One SiO$_2$/Si substrate spread by some WO$_3$ powders with another piece of blank SiO$_2$/Si substrate placed face-down above it were heated in a small one-end sealed quartz tube inside a big quartz tube by a tube furnace. Figure 1 b, c and d present optical microscopy, scanning electron microscopy (SEM) and atomic force microscopy (AFM) images of our CVD-grown WS$_2$ flakes on the covering substrate, respectively. It can be clearly seen that atomically thin flakes with perfect triangular shape and sharp edges are effectively formed on the typical 300 nm SiO$_2$/Si substrate through the CVD process. It is noticed that, though triangular WS$_2$ monolayers can be readily formed on both bottom pieces which contain WO$_3$ powders and covering pieces which are blank, the WS$_2$ monolayers grown on the covering substrates are much cleaner and appear as isolated perfect triangles (Fig. 1b and c). This could be due to the existence of WO$_3$ powders and highly concentrated reaction sources on the bottom substrates (see Supplementary Fig. S1). Own to the sensitivity to the local environment of the CVD growth, the distribution of dimension of the triangular crystals is quite wide, from a few microns to hundreds of microns while the majority of the flakes are more than 100 microns. The wide distribution of sizes is also observed in the previous reported CVD growth of MoS$_2$ and WS$_2$[26-27,30]. The zoom-in optical, SEM and AFM images show the clean surface of the WS$_2$ flakes which is very different with the MoS$_2$ and WS$_2$ thin layers grown by CVD previously[27,30] where impurity or small second, even multiple layers exist on top of the monolayers. The thickness of these triangular crystals is determined by the height profile of AFM and Raman spectroscopy (discussed later). Figure 1e - f present the TEM study of the WS$_2$ monolayers. The same as the optical, SEM and AFM images, a typical low-magnification TEM image (Fig. 1e) visualizes the perfect triangular shape and sharp edges of the as-grown WS$_2$ monolayers. The high resolution TEM (HRTEM) image (Fig. 1f) reveals the hexagonal ring lattice consisting of alternating tungsten atoms (dark dots) and sulfur (gray dots) atoms as schematically illustrated by the blue and yellow spheres. The TEM data also reveal the sharp edges at the micro-scale are perpendicular to the [100] crystalline direction. Though the micro-scale edges are along the zigzag direction, it should be noticed that our HRTEM image near the edge shows the edge is not atomically sharp, which ends up with the triangle-like features of a few nanometer (see Supplementary Fig. S2). Detailed study at the atomic level is needed. All data unambiguously show that the as-grown triangular WS$_2$ flakes are monolayer. The large dimension, perfect triangular shape, and clean surface indicate the WS$_2$ monolayer formed in this work could be a perfect candidate for studying fundamental knowledge and developing practical applications of 2D semiconductor.

**Lattice vibration and light emission of CVD-grown WS$_2$ monolayers.** Raman spectroscopy has been widely used to study 2D materials, such as the determination of numbers[31-35] and stacking sequence[36-38] of layers, the external field and molecular doping effects[39-40], and the internal and external strain[15-16,41-45]. Figure 2a shows Raman spectrum of the as-grown WS$_2$ monolayer over a frequency range of 80 – 650 cm$^{-1}$ at room temperature with the excitation laser of 532 nm. Fifteen Raman modes of WS$_2$ present, as labeled, including the first order modes of LA(M), LA(K), $E^1_{2g}(\Gamma)$, $E^1_{2g}(M)$ and $A_{1g}(\Gamma)$, the second order mode of 2LA(M), and some combinational modes. The frequency separation of 62.2 cm$^{-1}$ between the $E^1_{2g}(\Gamma)$ and $A_{1g}(\Gamma)$ [30] and the strong 2LA(M) mode at excitation laser of 532 nm caused by the double resonance scattering could be the spectral finger print of monolayer WS$_2$[46]. The Raman images (Fig. 2b) plotted by extracting the intensity and position of $A_{1g}(\Gamma)$ mode also clearly show the perfect triangular shape and uniform thickness of the as-grown WS$_2$ monolayers.

Figure 2d presents the fluorescence (FL) image of the as-grown WS$_2$ monolayers (their optical images are shown in Fig. 2c). It can be clearly seen that triangular WS$_2$ monolayers emit strong red luminescence resulted from the direct energy gap in such monolayers and the intensity of the light emission gradually decays from the edges to body centers of the triangles. These observations demonstrate that the light emission yield of the CVD $WS_2$ monolayers is extremely high and the FL image could be a quick and efficient tool to study such 2D direct energy gap semiconductors.

To probe the details of light emission from the CVD $WS_2$ monolayer, micro-photoluminescence (PL) spectroscopy is used and the PL spectra from the edge to the body center are plotted in Fig. 3a. Not only the intensities, but also the peak position and peak width of the PL vary from the edge to the center. As shown, blue-shift together with the quenching of the PL appears for the PL from the center relative to that of the edge. To exploit such non-uniformity cross the entire monolayer, we conducted PL mapping. Figure 3b shows the corresponding PL intensity image of the triangular $WS_2$ monolayer. The same as reflected in the FL image, the edges emit the strongest light, the strength of the emission gradually decays when moving towards the body center, and eventually becomes relatively "dark". The PL line scanning profiles of the location marked by the lines in Fig. 3b-d are shown in Fig. S3, which displays detailed evolution of PL intensity, peak width and peak position from the edge to the body. Though there is the reduction of light emission intensity from edges to body center, the overall PL, even from the body of $WS_2$ monolayer is comparable to the PL of the mechanical exfoliated $WS_2$ (MC-$WS_2$) monolayer, and is much stronger than that of the mechanical exfoliated $MoS_2$ (MC-$MoS_2$) monolayer as shown in Fig. 3f, 3h and 3j. It is noted that samples of even better quality can be formed by controlling the location and heating duration of the sulfur powder. For example, by heating the sulfur powders loaded at the upstream and outside of the furnace heating zone, $WS_2$ monolayers with uniform light emission which is as strong as the light emitting from MC-$WS_2$ monolayer, over the entire triangles were obtained (see Fig. S4). It is also noticed that a weak peak locating around 525 nm, around 0.4 eV higher than the dominant A exciton peak, presents in the PL spectrum of $WS_2$ monolayer under the excitation laser of 457 nm (see Fig. 3k), which is the so-called the B exciton emission[20-21]. The observation of B exciton peak, which was not seen in the CVD-grown $WS_2$ monolayers reported previously[5] further demonstrate the high optical quality of our large-scale $WS_2$ monolayers and the strong spin-orbit coupling. Another consequence, also great benefit of such strong spin-orbit coupling in $MS_2$ monolayer is the presence of valley-selective circular dichroism, which offers promising practical potential in valleytronics. Circularly polarized PL is an effective way to probe the valley dependence. Figure S5 presents the circular polarized PL spectra of mechanical exfoliated $MoS_2$ and the CVD-grown $WS_2$ monolayer under an excitation laser of 633 nm. The incident light is fixed as left-handed circularly polarized while the emitting lights of left- and right-handed circularly polarized are selectively collected. The net degree of the circular polarization is also plotted. As expected, an obvious valley-selective circular dichrism could be clearly observed in the CVD-grown $WS_2$ monolayers even at room temperature.

**Discussion**

As predicted and observed, the monolayer $WS_2$ possesses a direct energy gap of ~2 eV[5,19-21]. The strong PL peaks locating at 635 nm for the edges and 625 nm for the center are from the exciton emission. The blue-shift and the suppressing of PL at the center might be due to the existence of the structural and charged defects, such as S-vacancy as such kind of defect may cause the *n*-type doping to the $WS_2$ monolayer[47-48]. Sulfurization of $WO_3$ powder to produce $WS_2$ nanostructures has been widely adapted[30,49]. Recently, a modified two-step CVD process is used to grow nanometer and micrometer $WS_2$ monolayers on $SiO_2/Si$ substrate at 800 °C[30]. In this work, we developed a CVD process to grow hundreds of microns triangular single crystals of $WS_2$ monolayer on clean and blank 300 nm $SiO_2/Si$ substrates. In general, $WO_3$ powders on one $SiO_2/Si$ substrate together with another piece of covering blank $SiO_2/Si$ substrate were heated in a small diameter one-end sealed quartz tube inside a big diameter quartz tube by a tube furnace. To sulfurize the $WO_3$, sulfur powder was introduced into the upstream part of the quartz tube, where the temperature is around 200 °C – 250 °C. Different from the process reported previously[30], we used a one-end sealed inner quartz tube which was pre-cleaned by isopropyl alcohol (IPA) and deionized water inside a big quartz tube, which could effectively increase the concentration and pressure of vapor source for growth. During the whole process, argon gas was flowed with the flowing rate of 100 sccm. We propose the mechanism of such CVD process for growth of large-scale WS$_2$ monolayer as following. Firstly, flakes of WO$_y$S$_{2-y}$ were formed. Then, the further sulfurization produces triangular shape thick WS$_{2+x}$ flakes with the mixture of W$_{IV}$ and W$_{VI}$. As the apex of the triangles could be very active sites for nucleation, series of triangles formed and merged into a big triangle. With the continuous heating, the thick WS$_{2+x}$ flakes started expanding and thinning, and eventually WS$_2$ monolayers were fabricated. The formation of thin layers of WS$_2$ started at the center of the thick triangles since there has less overlapped small triangles and was exposed to sulfur for longest duration. It should be noticed that such center areas are also the most exposed regions after the sulfur source is exhausted during the final heating and cooling stages, which may cause the loss of the sulfur in the monolayers. The whole growth process can be well understood by the images presented in Fig. 4. Figures 4a–d are the schematic diagram illustrating the growth of WS$_2$ monolayer. Figures 4e-h show the optical images of the products at different growth stages, which usually can be simultaneously seen in one substrate, indicating the growth is very sensitive to the local environment. The formation of WS$_{2+z}$ at the thick parts are firmly proved by the energy dispersion spectroscopy (EDS) and time of flight secondary ion mass spectrometry (TOF-SIMS) images shown in Fig. S6. The EDS data indicate the ratios of W to S at the thick edge and thin body are 1:2.6 and 1:1.9, respectively. It is well known that W$_{VI}$ cannot be directly sulfurized by S unless some intermediate are formed[50]. Therefore, we think the transferring of W$_{VI}$ to W$_{IV}$ at the initial growth stage facilitates the growth of large-scale single crystals of WS$_2$ monolayer under a relatively relaxed condition. More systematic study of the growth mechanism is ongoing.

**Methods**

**Growth procedure.**

WS$_2$ monolayer flakes of fine triangular shape were grown by CVD method on well cleaned 300 nm SiO$_2$/Si substrates. In general, commercial WO$_3$ powders (> 99.5%, Sigma Aldrich) of 1 mg were spread on one piece of SiO$_2$/Si substrate and another piece of blank wafer was placed face-down above (2-3 mm) the bottom piece. They were then loaded into a small diameter quartz tube sealed at one-end which was pre-cleaned by IPA and deionized water. The small diameter quartz tube with 100 mg of sulfur powders (> 99.95%, Sigma Aldrich) located at the upstream part of it was introduced inside a big diameter quartz tube heated in a tube furnace. The furnace was heated firstly to 550 °C, then 750 °C with ramping rates of 20 °C/min and 3 °C/min, respectively. After the temperature was kept at 750 °C for 5 min, the furnace was cooled down to room temperature naturally. Ultra-high purity argon gas was flowed with the flowing rate of 100 sccm during the whole growth process.

**Optical and electronic microscopy study.**

The PL/Raman mappings/spectroscopies presented in this work were obtained by using a WITec Raman system with a piezocrystal controlled scanning stage. Lasers of 457 nm ($E_{laser}$ = 2.71 eV), 532 nm ($E_{laser}$ = 2.33 eV) and 633 nm ($E_{laser}$ = 1.96 eV) were used as the excitation sources. To avoid heating effect, the laser power at sample surface was controlled below 50 µW. The laser spot size is estimated to be 500 nm. A 100× objective lens with numerical aperture of 0.95 was used for normal PL/Raman measurement. Circularly polarized PL was performed by directing the linearly polarized laser passing through a quarter-wave plate to generate a circularly polarized light. The polarization of the incident light was set to be σ-. The polarization σ- or σ+ of backscattered PL was analyzed by a linear polarizer after the quarter-wave plate. A long-working distance 50× objective lens and a temperature controlled stage HFS600E from Linkam Scientific Instruments were used for circularly polarized PL measurement. The fluorescence images were obtained by an Olympus fluorescence microscope with a Mercury lamp as the excitation light source. The morphologies of as-grown $WS_2$ monolayer flakes were characterized by field emission scanning electron microscopy (FE-SEM, JEOL JSM-6700F). High-resolution TEM was carried out using an image aberration-corrected TEM system (FEI Titan 80-300). An acceleration voltage of 80 kV was chosen to achieve enough resolution while maintaining the structure of $WS_2$.

**EDS and TOF-SIMS.**

TOF-SIMS study of the as-grown $WS_2$ monolayer flakes on $SiO_2$/Si substrate was performed on TOF-SIMS [5] (ION-TOF GmbH) using 50 KeV $Bi_3^{++}$ as ion beam. The lateral resolution for SIMS imaging is 100 nm. EDS measurements were carried out using an Energy-dispersive X-ray analysis (EDX, Oxford Instruments) integrated with a Carl Zeiss Auriga system. 80 mm silicon-drift detector enables rapid determination of elemental compositions and acquisition of compositional maps.

**References**

1  Novoselov, K. S. *et al.* Electric field effect in atomically thin carbon films. *Science* **306**, 666-669 (2004).
2  van den Brink, J. GRAPHENE What lies between. *Nat. Mater.* **9**, 291-292 (2010).
3  Chhowalla, M. *et al.* The chemistry of two-dimensional layered transition metal dichalcogenide nanosheets. *Nat. Chem.* **5**, 263-275 (2013).
4  Wang, Q. H., Kalantar-Zadeh, K., Kis, A., Coleman, J. N. & Strano, M. S. Electronics and optoelectronics of two-dimensional transition metal dichalcogenides. *Nat. Nanotechnol.* **7**, 699-712 (2012).
5  Splendiani, A. *et al.* Emerging Photoluminescence in Monolayer $MoS_2$. *Nano Lett.* **10**, 1271-1275 (2010).
6  Mak, K. F., Lee, C., Hone, J., Shan, J. & Heinz, T. F. Atomically Thin $MoS_2$: A New Direct-Gap Semiconductor. *Phys. Rev. Lett.* **105**, 136805 (2010).
7  van der Zande, A. M. *et al.* Grains and grain boundaries in highly crystalline monolayer molybdenum disulphide. *Nat. Mater.* **12**, 554-561 (2013).
8  Radisavljevic, B., Radenovic, A., Brivio, J., Giacometti, V. & Kis, A. Single-layer $MoS_2$ transistors. *Nat. Nanotechnol.* **6**, 147-150 (2011).
9  Eda, G. *et al.* Photoluminescence from Chemically Exfoliated $MoS_2$. *Nano Lett.* **11**, 5111-5116 (2011).
10  Mak, K. F., He, K. L., Shan, J. & Heinz, T. F. Control of valley polarization in monolayer $MoS_2$ by optical helicity. *Nat. Nanotechnol.* **7**, 494-498 (2012).
11  Cao, T. *et al.* Valley-selective circular dichroism of monolayer molybdenum disulphide. *Nat. Commun.* **3**, 887 (2012).
12  Xiao, D., Liu, G. B., Feng, W. X., Xu, X. D. & Yao, W. Coupled Spin and Valley Physics in Monolayers of $MoS_2$ and Other Group-VI Dichalcogenides. *Phys. Rev. Lett.* **108**, 196802 (2012).
13  Sallen, G. *et al.* Robust optical emission polarization in $MoS_2$ monolayers through selective valley excitation. *Phys. Rev. B* **86**, 081301(R) (2012).
14  Feng, J., Qian, X. F., Huang, C. W. & Li, J. Strain-engineered artificial atom as a broad-spectrum solar energy funnel. *Nat. Photonics* **6**, 865-871 (2012).
15  Yanlong Wang, C. C., Caiyu Qiu, Ting Yu. Raman Spectroscopy Study of Lattice Vibration and Crystallographic Orientation of Monolayer $MoS_2$ under Uniaxial Strain. *Small*, DOI: 10.1002/smll.201202876 (2013).
16  Rice, C. *et al.* Raman-scattering measurements and first-principles calculations of strain-induced phonon shifts in monolayer $MoS_2$. *Phys. Rev. B* **87**, 081307(R) (2013).
17  Najmaei, S., Liu, Z., Ajayan, P. M. & Lou, J. Thermal effects on the characteristic Raman spectrum of molybdenum disulfide ($MoS_2$) of varying thicknesses. *Appl. Phys. Lett.* **100**, 013106 (2012).
18  Neville, R. A. & Evans, B. L. The Band Edge Excitons in 2H-$MoS_2$. *Phys. Status Solidi B* **73**, 597-606 (1976).
19  Ma, Y. D. *et al.* Electronic and magnetic properties of perfect, vacancy-doped, and nonmetal adsorbed $MoSe_2$, $MoTe_2$ and $WS_2$ monolayers. *Phys. Chem. Chem. Phys.* **13**, 15546-15553 (2011).

20  Frey, G. L., Tenne, R., Matthews, M. J., Dresselhaus, M. S. & Dresselhaus, G. Optical properties of $MS_2$ (M = Mo, W) inorganic fullerene-like and nanotube material optical absorption and resonance Raman measurements. *J. Mater. Res.* **13**, 2412-2417 (1998).
21  Ballif, C. *et al.* Preparation and characterization of highly oriented, photoconducting $WS_2$ thin films. *Appl. Phys. A-Mater.* **62**, 543-546 (1996).
22  Zeng, H. L. *et al.* Optical signature of symmetry variations and spin-valley coupling in atomically thin tungsten dichalcogenides. *Sci. Rep.* **3**, 1608 (2013).
23  Zeng, H. L., Dai, J. F., Yao, W., Xiao, D. & Cui, X. D. Valley polarization in $MoS_2$ monolayers by optical pumping. *Nat. Nanotechnol.* **7**, 490-493 (2012).
24  Britnell L, R. R., Eckmann A, Jalil R, Belle BD, Mishchenko A, Kim YJ, Gorbachev RV, Georgiou T, Morozov SV, Grigorenko AN, Geim AK, Casiraghi C, Castro Neto AH, Novoselov KS. Strong light-matter interactions in heterostructures of atomically thin films. *Science* **340**, 1311-1314 (2013).
25  Georgiou, T. *et al.* Vertical field-effect transistor based on graphene-$WS_2$ heterostructures for flexible and transparent electronics. *Nat. Nanotechnol.* **8**, 100-103 (2013).
26  Wu, S. F. *et al.* Vapor-solid growth of high optical quality $MoS_2$ monolayers with near-unity valley polarization. *ACS Nano* **7**, 2768-2772 (2013).
27  Sina Najmaei, Z. L., Wu Zhou, Xiaolong Zou, Gang Shi, Sidong Lei, Boris I. Yakobson, Juan-Carlos Idrobo, Pulickel M. Ajayan & Jun Lou. Vapour phase growth and grain boundary structure of molybdenum disulphide atomic layers. *Nat. Mater.* doi:10.1038/nmat3673 (2013).
28  Lee, Y. H. *et al.* Synthesis of large-area $MoS_2$ atomic layers with chemical vapor deposition. *Adv. Mater.* **24**, 2320-2325 (2012).
29  Liu, K. K. *et al.* Growth of large-area and highly crystalline $MoS_2$ thin layers on insulating substrates. *Nano Lett.* **12**, 1538-1544 (2012).
30  Humberto R. Gutiérrez, N. P.-L., Ana Laura Elías, Ayse Berkdemir, Bei Wang, Ruitao Lv, Florentino López-Urías, Vincent H. Crespi, Humberto Terrones, and Mauricio Terrones. Extraordinary room-temperature photoluminescence in triangular $WS_2$ monolayers. *Nano Lett.* DOI: 10.1021/nl3026357 (2013).
31  Ferrari, A. C. *et al.* Raman spectrum of graphene and graphene layers. *Phys. Rev. Lett.* **97**, 187401 (2006).
32  Li, S. L. *et al.* Quantitative Raman spectrum and reliable thickness identification for atomic layers on insulating substrates. *ACS Nano* **6**, 7381-7388 (2012).
33  Lee, C. *et al.* Anomalous Lattice Vibrations of Single- and Few-Layer $MoS_2$. *ACS Nano* **4**, 2695-2700 (2010).
34  Zhao, W. J. *et al.* Evolution of electronic structure in atomically thin sheets of $WS_2$ and $WSe_2$. *ACS Nano* **7**, 791-797 (2013).
35  Ni, Z. H. *et al.* Graphene thickness determination using reflection and contrast spectroscopy. *Nano Lett.* **7**, 2758-2763 (2007).
36  Lui, C. H. *et al.* Imaging Stacking Order in Few-Layer Graphene. *Nano Lett* **11**, 164-169 (2011).
37  Cong, C. X., Yu, T., Saito, R., Dresselhaus, G. F. & Dresselhaus, M. S. Second-order overtone and combination Raman modes of graphene layers in the range of 1690-2150 cm$^{-1}$. *ACS Nano* **5**, 1600-1605 (2011).
38  Cong, C. X., Yu, T. & Wang, H. M. Raman study on the G mode of graphene for determination of edge orientation. *ACS Nano* **4**, 3175-3180 (2010).
39  Chakraborty, B. *et al.* Symmetry-dependent phonon renormalization in monolayer $MoS_2$ transistor. *Phys. Rev. B* **85**, 161403(R) (2012).
40  Tongay, S. *et al.* Broad-range modulation of light emission in two-dimensional semiconductors by molecular physisorption gating. *Nano Lett.* **13**, 2831-2836 (2013).
41  Yu, T. *et al.* Raman mapping investigation of graphene on transparent flexible substrate: The strain effect. *J. Phys. Chem. C* **112**, 12602-12605 (2008).

42  Huang, M. Y. *et al.* Phonon softening and crystallographic orientation of strained graphene studied by Raman spectroscopy. *P. Natl. Acad. Sci. USA* **106**, 7304-7308 (2009).
43  Huang, C. W. *et al.* Revealing anisotropic strain in exfoliated graphene by polarized Raman spectroscopy. *Nanoscale* 10.1039/c1033nr00123g (2013).
44  Mohiuddin, T. M. G. *et al.* Uniaxial strain in graphene by Raman spectroscopy: G peak splitting, Gruneisen parameters, and sample orientation. *Phys. Rev. B* **79**, 205433 (2009).
45  Ni, Z. H. *et al.* Uniaxial Strain on Graphene: Raman Spectroscopy Study and Band-Gap Opening. *ACS Nano* **2**, 2301-2305 (2008).
46  Berkdemir, A. *et al.* Identification of individual and few layers of $WS_2$ using Raman Spectroscopy. *Sci. Rep.* **3**, 1755 (2013).
47  Zhou, W. *et al.* Intrinsic structural defects in monolayer molybdenum disulfide. *Nano Lett.* **13**, 2615-2622 (2013).
48  Mak, K. F. *et al.* Tightly bound trions in monolayer $MoS_2$. *Nat. Mater.* **12**, 207-211 (2013).
49  Tenne, R., Margulis, L., Genut, M. & Hodes, G. Polyhedral and Cylindrical Structures of Tungsten Disulfide. *Nature* **360**, 444-446 (1992).
50  van der Vlies, A. J., Kishan, G., Niemantsverdriet, J. W., Prins, R. & Weber, T. Basic reaction steps in the sulfidation of crystalline tungsten oxides. *J. Phys. Chem. B* **106**, 3449-3457 (2002).

**Acknowledgements**

This work is supported by the Singapore National Research Foundation under NRF RF Award No. NRFRF2010-07 and MOE Tier 2 MOE2012-T2-2-049. C.C.X. and T.Y. thank the help of WinTech for the TOF-SIMS analysis. X.W and L.T.S thank the support of Jiangsu Province Funds (No. BK2012024) and Chinese postdoctoral funding (No. 2012M520053).

**Author contribution**

C.C.X and T.Y. initialled the project, conceived and designed the CVD growth; C.C.X., B.C.C. and T.Y. performed the optical measurements; X.W. and L.T.S. conducted TEM characterization. J.Z.S, C.Y.Q. and N.P. helped on the sample preparation. C.C.X. and T.Y. collected the data and co-wrote the paper. All authors discussed the results and commented on the manuscript.

**Additional information**

Competing financial interests: The authors declare no competing financial interests.

**Figure legends**

**Figure 1 │ Large-scale single crystalline monolayer $WS_2$ growth and characterization. (a)** Schematic diagram of the CVD system used for the growth of $WS_2$. **(b)** Optical image of as-grown $WS_2$

on SiO$_2$ (300 nm)/Si substrate. Inset shows a triangular monolayer WS$_2$. **(c)** SEM image of the area shown in (a). **(d)** AFM image of a monolayer WS$_2$. The height profile demonstrates the as-grown WS$_2$ flakes are monolayers. **(e, f)** Low- and high-magnification TEM images of triangular WS$_2$ monolayer, respectively. Optical, SEM, AFM and TEM images show the perfect triangular shape and sharp edges (at micron scale) of the as-grown WS$_2$ monolayers.

**Figure 2 │ Lattice vibrations and light emission of WS$_2$ monolayers. (a)** Raman spectrum of WS$_2$ monolayer at room temperature. **(b)** Raman image of a triangular single crystal of WS$_2$ monolayer constructed by plotting A$_{1g}$ mode intensity (upper panel) and peak position (lower panel). **(c)** Optical image of as-grown WS$_2$ on SiO$_2$ (300 nm)/Si substrate. **(d)** Fluorescence images of the as-grown WS$_2$ flakes shown in the optical image of (c).

**Figure 3 │ Photoluminescence (PL) and high quantum yield of light emission of CVD grown WS$_2$ monolayers.** (a) PL spectra of the corresponding (colors) points shown in (b). The inset of (a) shows the optical image of the monolayer WS$_2$ flake studied here. (b) – (d) PL images of the peak integrated intensity, position and width, respectively. The lines indicate the locations for the PL line scanning investigation (see Supplementary Fig. S3). (e, g, i) Optical images of mechanically exfoliated MoS$_2$ (MC-MoS$_2$) and WS$_2$ (MC-WS$_2$) monolayer, and CVD-WS$_2$ monolayer. (f, h, j) The corresponding fluorescence (FL) images of the samples shown in (e, g and i). The FL images are taken under the identical conditions for MoS$_2$ and WS$_2$ monolayer flakes. The bright FL image indicates the ultrahigh quantum yield of direct bang gap light emission from our CVD-grown WS$_2$ monolayers. (k) PL spectrum of CVD WS$_2$ monolayer excited by 457 nm laser. The B exciton peak can be clearly seen in the inset.

**Figure 4 │ Growth mechanism of large-scale triangular single crystals of WS$_2$ monolayers by CVD process. (a – d)** Schematic diagram illustrates the growth of WS$_2$ monolayers. **(e – h)** Optical images of the flakes at the different growth stages. Note: the flakes are not the same piece.

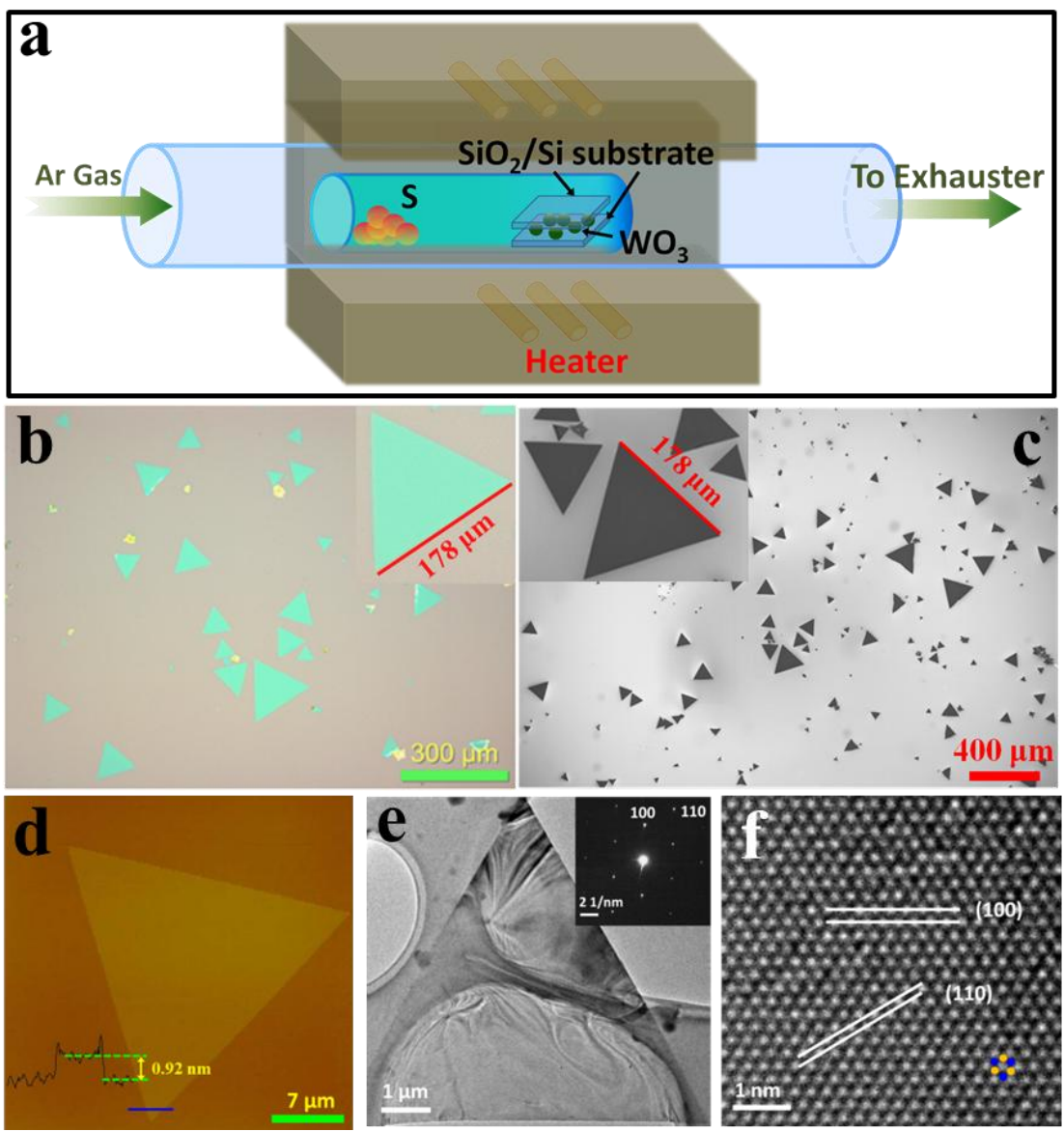

Figure 1

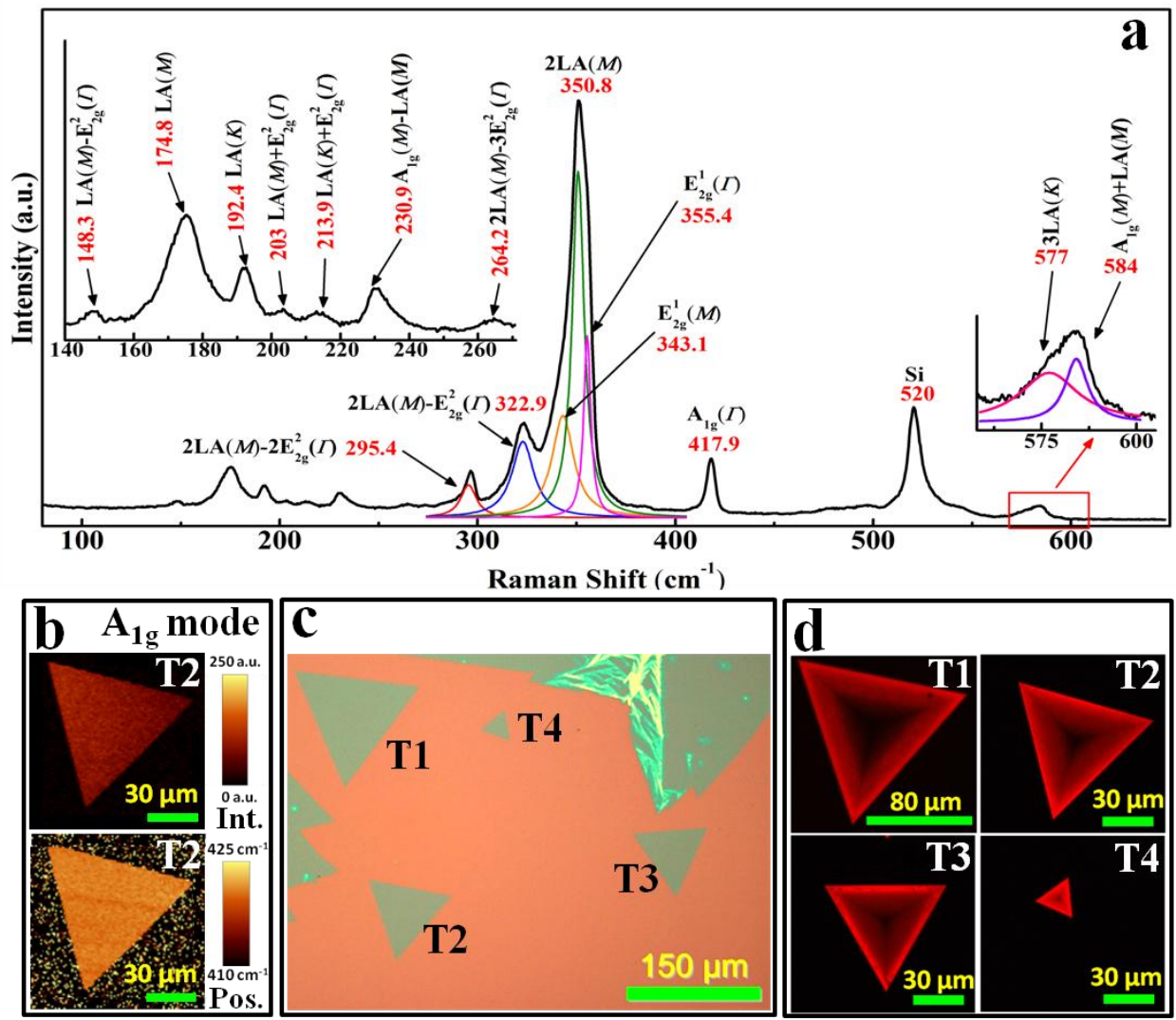

Figure 2

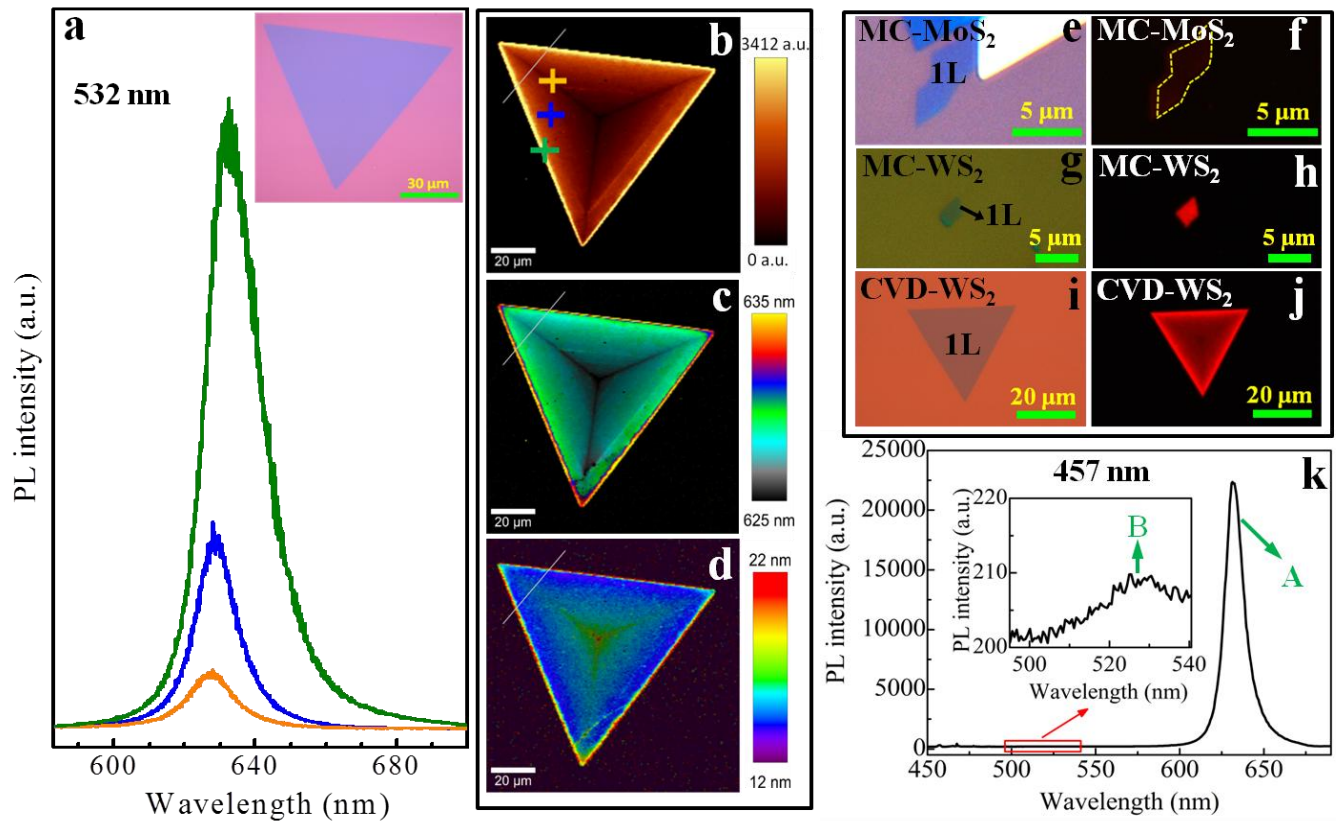

Figure 3

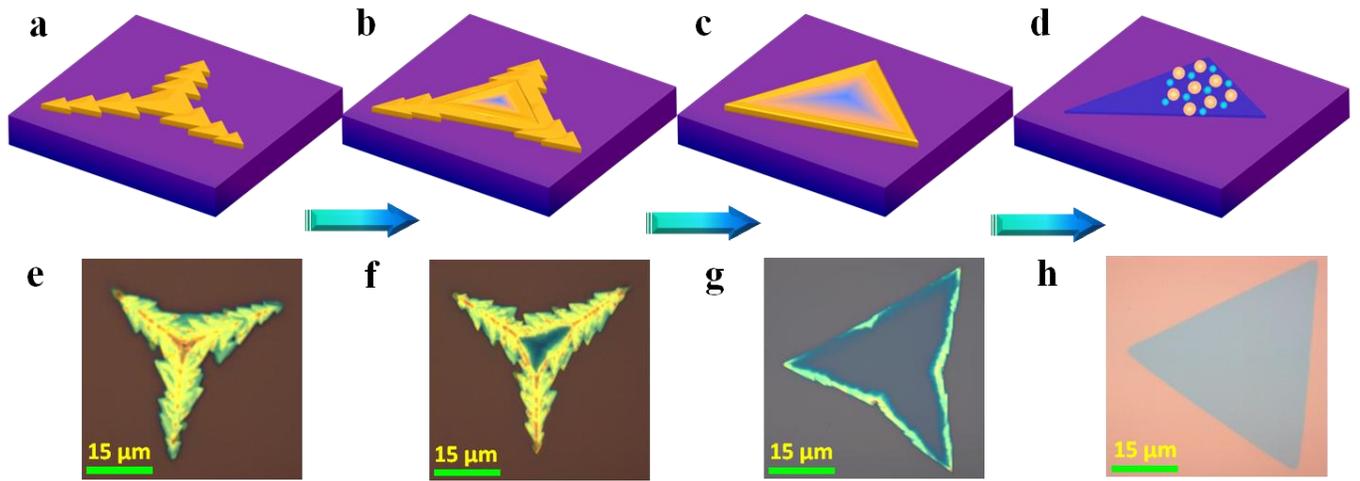

Figure 4

# Supplementary Information

**Synthesis and optical properties of large-scale single-crystalline two-dimensional semiconductor $WS_2$ monolayer from chemical vapor deposition**

Chunxiao Cong[1], Jingzhi Shang[1], Xing Wu[2], Bingchen Cao[1], Namphung Peimyoo[1], Caiyu Qiu[1], Litao Sun[2], and Ting Yu[1, 3, 4, *]

[1] Division of Physics and Applied Physics, School of Physical and Mathematical Sciences, Nanyang Technological University, 637371, Singapore

[2] SEU-FEI Nano-Pico Center, Key Laboratory of MEMS of Ministry of Education, School of Electrical Science and Engineering, Southeast University, Nanjing 210096, China

[3] Department of Physics, Faculty of Science, National University of Singapore, 117542, Singapore

[4] Graphene Research Center, National University of Singapore, 117546, Singapore

*Address correspondence to yuting@ntu.edu.sg

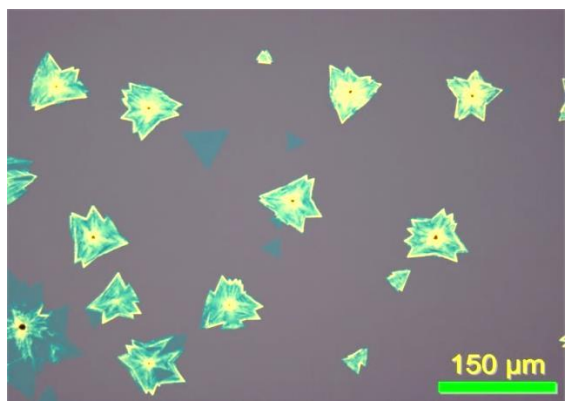

**Supplementary Figure S1: Morphology of WS$_2$ flakes grown on bottom substrate.** Optical image of the as-grown WS$_2$ flakes. Dark dots at the centre of flakes are the initial locations of WO$_3$ particles. With the abundant sources, big pieces of multiple domain flakes with thick parts are formed. From the figure, it can be seen that the formation of multiple domain flakes occurs at the initial stage of growth rather than by merging of the single triangular monolayers when they meet up with each other.

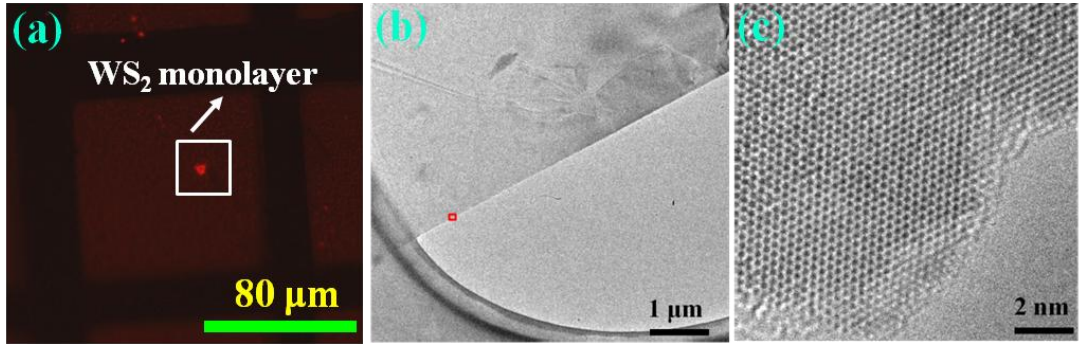

**Supplementary Figure S2: FL image of WS$_2$ monolayer transferred onto TEM grid and HRTEM of the edge of WS$_2$ monolayer.** (a) The fluorescence (FL) image of WS$_2$ monolayer transferred onto TEM grid. (b) Typical low-magnification TEM image of WS$_2$ monolayer, show the sharp edge at micro-scale level. (b) HRTEM image of the selected area in (b). A nano-scale rough edge is present. The same as the observation shown in Fig. 1f, the monolayer shows high quality of single crystalline nature and the micro-scale sharp edge is along the zigzag direction.

The change of contrast at the very edge region could be due to the ripples. 2D membranes embedded in a 3D space have a tendency to be crumpled[1]. These fluctuations can, induce bending and stretching into the WS$_2$ membrane, thus exhibit strong height differences (contrast).

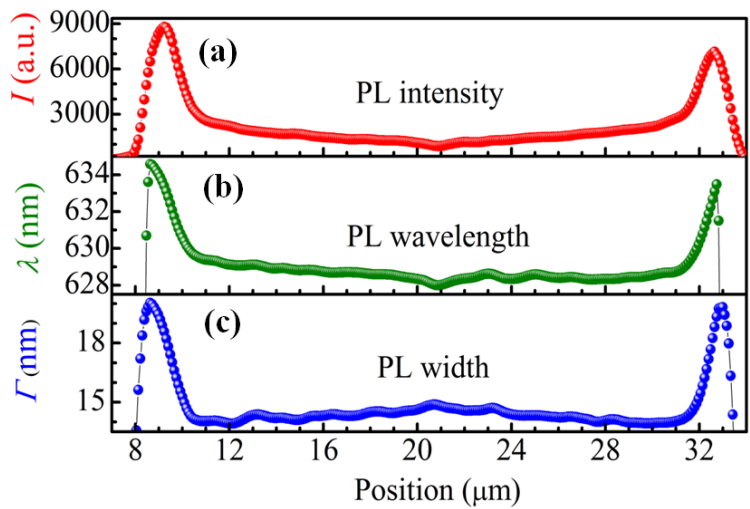

**Supplementary Figure S3: PL line scanning characterization of $WS_2$ monolayer.**

Line scan profile of PL spectral features of (a) integrated intensity, (b) peak position (or wavelength) and (c) peak width crossing the $WS_2$ monolayer marked by lines in Fig. 3b-d. As the PL mappings and the spectra (Fig. 3) shown, the PL at the body centre of the monolayers is suppressed and shifts towards short wavelength, which could be caused by the existence of structural and charged defects such as sulfur vacancy there.

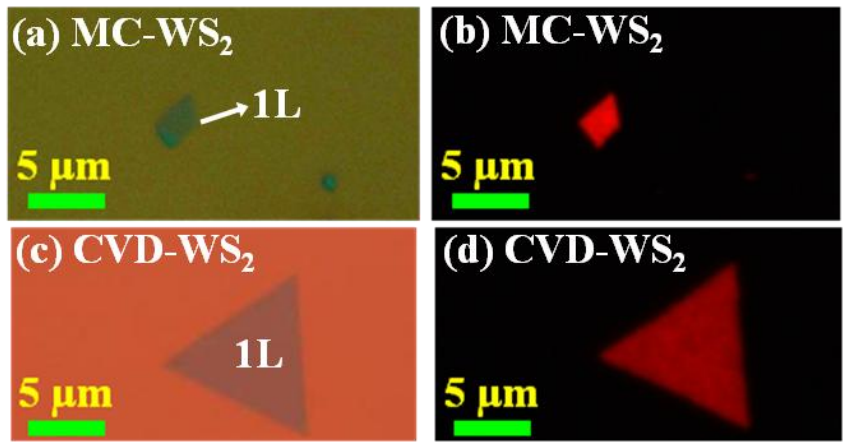

**Supplementary Figure S4: Optical and FL images of mechanically exfoliated WS$_2$ monolayer and CVD-WS$_2$ monolayer.** (a, c) Optical images of mechanically exfoliated WS$_2$ (MC-WS$_2$) monolayer and CVD-WS$_2$ monolayer. (b, d) The corresponding fluorescence (FL) images of the samples shown in (a, c). The FL images are taken under the identical conditions for MC-WS$_2$ and CVD-WS$_2$ monolayer flakes.

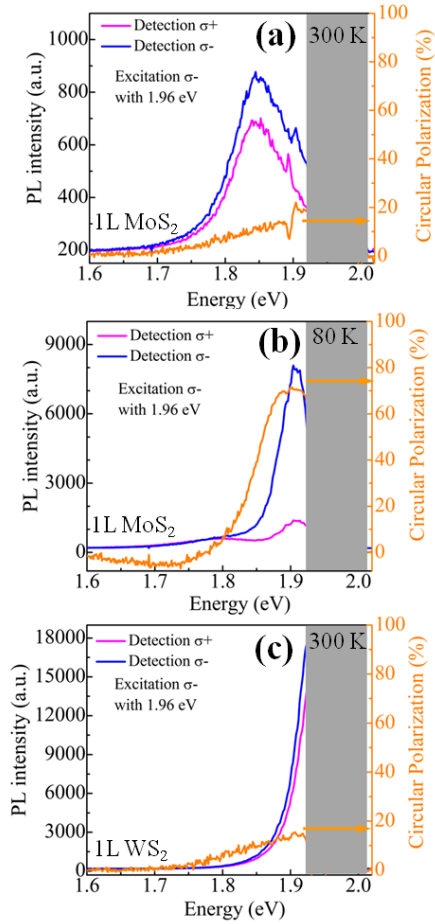

**Supplementary Figure S5: Valley-dependent circular dichroism of mechanical exfoliated MoS$_2$ and the CVD-grown WS$_2$ monolayers.** Circularly polarized PL spectra excited by 633 nm laser of mechanical exfoliated MoS$_2$ monolayer at (a) 300 K, (b) 80 K, and (c) CVD-grown WS$_2$ monolayer at 300 K. The circular polarization of the PL spectra is also plotted. The gray band indicates the spectra blocked by the laser filter. It is also noticed that such valley dependency becomes maximum near the most intensive PL region. As only the low energy-side tails of PL spectra of WS$_2$ monolayer can be obtained, a higher circular polarization is expected when comparing the peak maxima. This is under ongoing investigation.

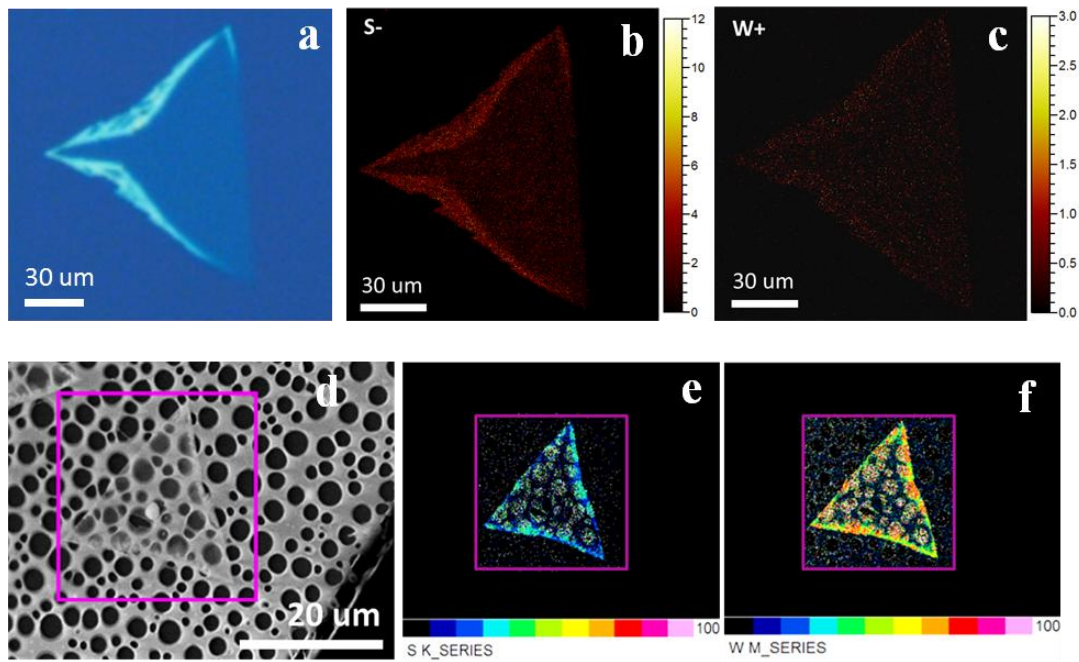

**Supplementary Figure S6: Elemental analysis of WS$_2$ flake at the transition stage of growth.** (a) Optical image of WS$_2$. (b, c) Time-of-flight secondary ion mass spectrometry (TOF-SMIS) images of negative sulfur ion and positive tungsten ion from the flake shown in (a). (d) SEM image of the WS$_2$ flake at the same growth stage of the sample in (a) which was transferred onto a TEM grid. (e, f) Scanning electron microscopy – X-ray energy dispersive spectroscopy (SEM-EDS) images of S-K and W-M from the sample in (d). The TOF-SIMS image of S- reveals, for the top 1-2 atomic layers, the thick parts at the edges contain more S than the thin region does. The EDS mappings also show sulfur to tungsten ratio is higher at the thicker parts.

## References

1. A. Fasolino, J. H. Los & M. I. Katsnelson, "Intrinsic ripples in graphene", Nature Materials 6, 858 - 861 (2007).